\documentclass{JHEP}
\usepackage{epsfig}
\preprint{ {\tt hep-th/0012089} }
\newcommand{\be}{\begin{equation}}
\newcommand{\ee}{\end{equation}}
\newcommand{\bea}{\begin{eqnarray}}
\newcommand{\eea}{\end{eqnarray}}
\newcommand{\eq}[1]{(\ref{#1})}
\newcommand{\del}{\partial}

\title{ Tachyon condensation using the disc partition function }
\author{Justin R. David \\ Department of Physics, University of
California\\Santa Barbara, CA 93106, USA.\\
\email{justin@vulcan.physics.ucsb.edu} }
\abstract{ 
It has been recently proposed that the background independent 
open superstring field
theory action is given by the disc partition function with all
possible open string 
operators inserted at the boundary of the disc. We use this
proposal to study tachyon condensation in the D0-D2 system. 
We evaluate the disc partition function for the D0-D2 system
in presence of a large Neveu-Schwarz B-field using perturbation
theory. This perturbative expansion of the disc partition function
makes sense as the boundary tachyon operator for the large
Neveu-Schwarz B-field is almost marginal.
We find that the mass defect for the formation of the D0-D2
bound state agrees exactly with the expected result 
in the large B-field limit.  }

\begin{document}
\section{Introduction}

Recent works have addressed the study of 
open string tachyon condensation from
various point of view. In this paper we focus on the string field
theoretic approach to study tachyon condensation. The 
open string tachyon in bosonic string theory signals the instablity of
the D25-brane to decay to the vacuum. There exists a 
a stationary point in the tachyon potential \cite{kossam,sen}. 
The value of the tachyon potential at this stationary point 
should agree with the tension
of the unstable brane. This has been verified to a remarkable degree
of accuracy in cubic string field theory using level truncation
\cite{kossam, senzwi, wati, moewati,raszwi}. 
The physics of the solitons in
the tachyon potential also seem to be reproduced in the level
truncation approximation scheme \cite{harkra,kjmt,msz}.
More recently, the exact tachyon potential has
been calculated using the 
background independent open string field theory formulated by
\cite{witten, shat}.
The stationary point of the tachyon potential from background
independent open string field theory agrees precisely with the tension
of the D-brane \cite{gersha,kmm,ghosen}. Condensation to lower
dimensional branes also reproduced their expected tensions.

There is a similar story for the case of unstable branes and D-brane
anti-D-brane systems of superstring field theory. For these systems
the tachyon potential has been evaluated using level truncation in the
superstring theory formulated by Berkovits \cite{berk}.
The stationary point of
this potential agrees with the tension of the branes to a high degree
of accuracy \cite{berk2,bsz,smerae,iqbnaq}. 
It has been recently proposed in \cite{kmm2} that the
The background independent superstring field theory action 
is the disc
partition function with all possible operators inserted on the
boundary of the disc. Precisely the boundary string field theory
action is given by
\be
S[\lambda_i] = Z[\lambda_i],
\ee
where $Z[\lambda_i]$ is the disc partition function of the world sheet
conformal field theory. $\lambda_i$ are the couplings of the various
boundary operators. This is a generalization of the proposal of
\cite{andtse, tse1} which restricted the operators to those
corresponding to the massless modes (See also \cite{tse2}).
Using this proposal tachyon condensation on
unstable D-branes of type II sting theory was studied in \cite{kmm2}.
The tachyon potential found had a minimum which agreed with the tension
of the unstable D-brane. Condensation to lower dimensional branes
reproduced their expected tensions \footnote{Exact results for tachyon
condensation has also been obtain by introducing noncommutativity
\cite{gms,kmr,hklm,jmw,agms}. Recently Matrix theory has been used to
study tachyon condensation \cite{krs,mli,gase}}.

There are open string tachyons in other systems. Two examples of such 
systems are The D0-D4 system with a Neveu-Schwarz B-field in the
spatial directions of the D4-brane \cite{seiwit}.
and the D0-D2 system. The tachyon potential for the D0-D2 system was
studied in first quantized string theory in \cite{gns}. It is
interesting to study tachyon condensation in these systems for the
following reasons. These unstable systems decay to BPS states unlike
the unstable D-branes which decay to the vacuum. The tachyon potential
for these systems lies outside the universality class of the unstable
D-branes \cite{jus}. In fact they have a parameter 
, the B-field which offers a controlled
expansion in the study of the tachyon potential. This parameter allows
the interpolation between these systems and 
the D-brane anti-D-brane system. 
Study of tachyon condensation in these systems offer tests to
proposals of superstring field theory action. This is important
because the formulation of superstring field theory is not complete.

In this paper we will use the 
the disc partition function to study
tachyon condensation in the D0-D2 system. In general the disc
partition function with all possible boundary operators is not well
defined. For unstable D-branes it was possible to evaluate the disc
partition function exactly as the tachyon boundary operator was linear
and the theory was free. It was also argued that this operator does
not mix with any others in renormalization group flow. For the D0-D2
system the tachyon operator is a twist operator. We do not have the
facility of a free theory. But we have the facility of making this
operator almost marginal by introducing a B-field in the spatial
direction of the D2-brane. For large values of the B-field the tachyon
operator becomes almost marginal.
Then we can set up a 
well defined perturbative expansion
of the disc partition function in the coupling $\lambda$
\footnote{The disc partition function with the B-field turned on has
been evaluated in \cite{lc} for the bosonic case.}. We show that
the leading order term in the 
mass defect obtained from the disc partition
function agrees exactly with the expected mass defect from the BPS
formula. This lends support to the proposal that the disc partition
function is the background independent superstring field theory
action.

The organization of this paper is as follows. In section 2 we review 
the formulation of how to 
introduce 
boundary interaction in the disc partition function
in a manifestly world sheet supersymmetric manner
\cite{wit,hkm,kmm2}. These
interactions correspond to boundary tachyonic interaction.
We also review the point splitting regularization introduced by
\cite{kmm2}. In section 3 we introduce the D0-D2 system with the
B-field and construct the boundary tachyon operators. We evaluate the
various correlation functions of these operators required to evaluate
the tachyon potential. In section 4 we use all the ingredients to
evaluate the disc partition function till the quartic order in
couplings. In section 5 we compare the mass defect obtained from the
disc partition function to the leading order from the BPS formula. We
obtain exact agreement.
Section 6 contains our conclusions. The appendices contains details
regarding the tachyon operator and the various correlation functions.

\section{The disc partition function in perturbation theory.}
The disc partition function for the world sheet 
of the Neveu-Schwarz superstring is given by
\be
Z[\lambda_i] = \int [d \psi^\mu][d X^\mu] e^{-S_{\rm{Bulk}}
-S_{\rm{Boundary}} }
\ee
Where
\be
S_{\rm{Bulk}} = \frac{1}{4\pi} \int d^2 z \left( 
\del X^\mu \bar{\del} X_\mu
+ \psi^\mu\bar{\del}\psi_\mu
+ \tilde{\psi}^\mu\del\tilde{\psi}_\mu \right)
\ee
Here $\mu$ runs from $0\ldots 9$ and the signature of the world sheet
and the space time is Euclidean. We have set $\alpha'=2$, 
the integral in $S_{\rm{Bulk}}$ is over a disc of radius $1$.
The disc partition function depends on $\lambda_i$ which stands for
the various couplings in the boundary interaction.
We introduce the boundary interaction in $S_{\rm{Boundary}}$ 
preserving $N=1$ supersmmetry. The boundary 
superspace coordinates are
$(\tau, \theta)$ with $\tau\in [-\pi, \pi]$ and $\theta$ the boundary
Grassman coordinate.
The boundary action in this superfield notation is given by
\be
\label{bound}
S_{\rm{Boundary}} = -\int \frac{d\tau}{2\pi} d\theta \left(
\Gamma D\Gamma + (O_1 + \theta O_2) \Gamma \right) ,
\ee
here
\be
D= \del_\theta + \theta \del_\tau \;\;\;\;\;
\Gamma = \eta + \theta F.
\ee
$\Gamma$ is the fermionic degree of freedom living on the
boundary. $O_1$ and $O_2$ are the components of the superfield
corresponding to the tachyonic boundary interaction. 
The integral in \eq{bound} is over the boundary of the disc and runs
from $-\pi$ to $\pi$.
Eliminating $\Gamma$ by
using its equation of motion and integration over $\theta$ we obtain
the following boundary action.
\be
S_{\rm{Boundary}} = \frac{1}{4}\int \frac{d\tau}{2\pi} \left(
O_1^2 (\tau) + O_2(\tau) \int \frac{1}{2}\Theta(\tau -\tau')
O_2(\tau') d\tau' \right)
\ee
where $\Theta (\tau) =1$ for $\tau>0$ and $\Theta (\tau) = -1$ for
$\tau<0$. We also have
\be
\frac{d}{d\tau} \Theta(\tau) = 2\delta(\tau)
\ee
Though the boundary interaction is non-local it is well defined in
the Neveu-Schwarz sector as $\eta$ and $\psi$ do not have zero modes
in this sector.
We formulate the perturbative expansion of the disc partition function
by simply expanding in powers of the boundary interaction. We have
the following expansion. 
\bea
\label{pert}
\frac{Z[\lambda_i]}{Z[0]} 
&=& 1 - \langle S_{\rm{Boundary}}\rangle + \frac{1}{2}
\langle (S_{\rm{Boundary}} )^2\rangle + \cdots \\ \nonumber
&=& 1 + S_2 + S_4 + \cdots
\eea
We will justify this expansion later for  boundary
interactions which are almost marginal. 
We will be interested 
only
till the quartic term in the expansion of the disc partition function
in \eq{pert}.

To evaluate each term in the expansion in \eq{pert} we have to use a
renormalization prescription. We use a point splitting renormalization
prescription introduced by \cite{kmm2}. 
For the term $S_2$  in \eq{pert} it is given by
\be
S_2= \lim_{\epsilon \rightarrow 0}  
-\frac{1}{4}\left(\int \frac{d\tau}{2\pi}\langle
O_1(\tau) O_1(\tau-\epsilon) + O_2(\tau) \int \frac{1}{2} \Theta
(\tau -\tau'-\epsilon) O_2 (\tau') d\tau' \rangle \right)
\ee
It is easy to see that this renormalization prescription preserves
world sheet supersymmetry as $S_2$ is
$\frac{1}{2}\int d\tau d\theta \langle (O_1 (\tau) + \theta
O_2(\tau))\Gamma(\tau-\epsilon, \theta)\rangle $. 
Carrying over this prescription to the 
quartic term $S_4$ we get
\be
\label{limit}
S_4 = \frac{1}{8}
\lim_{\epsilon_1\rightarrow 0, \epsilon_2\rightarrow 0}
\int d\tau_1 d\tau_2 d\theta_1 d\theta_2
\langle (O_1 + \theta_1 O_2) \Gamma(\tau_1 -\epsilon_1, \theta_1)
(O_1 + \theta_2 O_2) \Gamma(\tau_1 -\epsilon_2, \theta_2) \rangle
\ee
To simplify the calculations we take the limit 
in \eq{limit} along the $45^\circ$
line in the $\epsilon_1, \epsilon_2$ plane. Therefore to evaluate the
quartic term in the expansion of the disc partition function we use
the prescription
\be
S_4 = \frac{1}{8}\lim_{\epsilon\rightarrow 0}
\int d\tau_1 d\tau_2 d\theta_1 d\theta_2 \langle
(O_1 + \theta_1 O_2) \Gamma(\tau_1 -\epsilon, \theta_1)
(O_1 + \theta_2 O_2) \Gamma(\tau_1 -\epsilon, \theta_2) \rangle
\ee

\section{The tachyon in the D0-D2 system}

In this section we review the D0-D2 system 
with a Neveu-Schwarz B-field along the spatial directions of the
D2-brane. See \cite{citmm} for a discussion of the D$p$-D$p'$ system
with the Neveu-Schwarz B-field.
We then construct the tachyon operators $O_1$ and $O_2$ for
this system.

\subsection{The D0-D2 system}
Consider a single D0-brane and a single D2-brane of type IIA string
theory in ten dimensions configured as follows. The D2-brane is
extended along the directions $x^1$ and $x^2$. The open string
spectrum consists of excitations of strings joining the D0-brane 
with itself and the D2-brane with itself. Then there are excitations
of the open string joining the D0-brane with the D2-brane. We denote
the open strings joining the D0-brane with the D2-brane as the $(0,2)$
strings. The lowest excitation of the 
$(0,2)$ strings is a tachyon. Let the string world sheet coordinates
of the $(0,2)$ string be $X^\mu (\sigma^0, \sigma^1). \mu$ runs form
$0, \ldots, 9$, and $\sigma^1$ lies between $0$ and $\pi$. We work
with Euclidean world sheet signature. Now turn on a constant B-field
along the spatial directions of the D2-brane. This changes the mass
of the $(0,2)$ tachyon. The B-field is given by
\be
B_{ij} = \frac{1}{4\pi}\left(
\begin{array}{cc}
0 & b \\
-b & 0
\end{array}
\right),
\ee
where $i, j \in \{1, 2\}$. We choose the metric $g_{ij} = \delta_{ij}$.
The boundary conditions of the world sheet coordinates with these
moduli turned on is given by

\bea
\left.
\partial_{\sigma^1} X^i + 4\pi i  B_{ij}\partial_{\sigma^0} 
X^j \right|_{\sigma^1 = \pi} = 0 \\ \nonumber
\partial_{\sigma^0} X^i |_{\sigma^1 = 0} = 0 \\ \nonumber
\partial_{\sigma^0} X^a |_{\sigma^1 = 0 , \sigma^1 = \pi} = 0 
\hbox{ where } a =  3\ldots 9 \\ \nonumber
\partial_{\sigma^1} X^0 |_{\sigma^1 = 0, \sigma^1 = \pi} = 0
\eea
The non-trivial mode expansions arise for the world sheet coordinates-ordinates
$X^i$, It is convenient to define coordinates-ordinates 
\be
X^+ = X^1+ iX^2 \;\;\;\; X^- = X^1- i X^2 
\ee
The mode expansions of $X^+$ and $X^-$ are given by
\bea
\label{bosemode1}
X^+ &=& i\sqrt{2} 
\sum_n \left[  \frac{\alpha^+_{n-\nu}}{n-\nu} e^{-(n-\nu)(\sigma^0 +
i\sigma^1)} - \frac{\alpha^+_{n-\nu}}{n-\nu} e^{-(n-\nu) 
(\sigma^0 - i\sigma^1)} \right] \\ \nonumber
X^- &=& i\sqrt{2} 
\sum_n \left[  \frac{\alpha^-_{n+\nu}}{n+\nu} e^{-(n+\nu)(\sigma^0 +
i\sigma^1)} - \frac{\alpha^-_{n+\nu}}{n+\nu} e^{-(n+\nu)
(\sigma^0 - i\sigma^1)} \right]
\eea
Where 
\be
\label{phase1}
e^{2\pi i \nu} = -\frac{1+ ib}{ 1-ib}, \;\;\; 0\leq\nu<1 
\ee
The only non zero commutation relations are 
\be
[\alpha^{+}_{n-\nu}, \alpha^{-}_{m +\nu} ] = (n-\nu) \delta(n +m)
\ee

The mode expansions of world sheet superpartners of the bosonic fields
are fixed by supersymmetry. The mode expansions of $\psi^+$ and
$\bar{\psi}^+$ of $X^+$  is given by
\bea
\label{mode1}
\psi^+ &=& -i\sqrt{2} \sum_n
\psi^+_{n+1/2-\nu}e^{-(n+1/2 -\nu)(\sigma^0
+ i \sigma^1)} \\ \nonumber
\bar{\psi}^+ &=& i\sqrt{2} \sum_n
\psi^+_{n+1/2-\nu}e^{-(n+1/2 -\nu)(\sigma^0
- i \sigma^1)}
\eea
The mode expansions of the superpartners of $X^-$ are given by
\bea
\label{mode2}
\psi^- &=& -i\sqrt{2} \sum_n
\psi^+_{n+1/2+\nu}e^{-(n+1/2 +\nu)(\sigma^0
+ i \sigma^1)} \\ \nonumber
\bar{\psi}^+ &=& i\sqrt{2} \sum_n
\psi^+_{n+1/2+\nu}e^{-(n+1/2 +\nu)(\sigma^0
- i \sigma^1)}
\eea
We have written the mode expansions in the Neveu-Schwarz sector. The
only non-zero anti-commutation rules are given by
\be
\label{anticom}
\{ \psi^+_{n+1/2 -\nu} , \psi^-_{m+1/2+\nu}  \} = \delta(m +n)
\ee

The (mass)$^2$ of tachyon in the 
$(0,2)$ strings is determined from the zero
point energy. The zero point energy for two of the 
lowest states in the Neveu-Schwarz sector given by
\be
E_0 = -\frac{1}{2} + \frac{\nu}{2} \;\;\; E_+ = -\frac{\nu}{2}
\ee
The state with energy $E_0$ is retained by the GSO projection.
We see that the (mass)$^2$ of the 
tachyon decreases as $b\rightarrow\infty$. As $b\rightarrow\infty$ we
have $\nu\rightarrow 1- 1/(\pi b)$. Thus
the zero point energy in the limit of large $b$ is given by
$E_0 = -\frac{1}{\pi b}$. For $b\rightarrow -\infty$ the mode
expansions in \eq{bosemode1}, \eq{mode1} and \eq{mode2} reduce to that
of Dirichlet-Dirichlet boundary conditions. The induced 
D0-brane charge on the D2-brane is proportional to $b$. Therefore for
$b\rightarrow -\infty$ the system reduces to the D0-brane
anti-D0-brane system. In fact in this limit the zero point energy $E_0
= -1/2$ which is that of the tachyon of the D0-brane anti-D0-brane
system.

\subsection{The calculation of the mass defect}

The tachyon in the (0,2) strings signal the instability to form a
D0-D2 bound state. In this section we will use BPS mass formulae to
evaluate the mass defect in the formation of the bound state.
The mass of the D0-brane is given by \footnote{ Note that we have used
$\alpha' =2$. }
\be
M_{D0} = \frac{1}{g\sqrt{2}}
\ee
The Mass of the D2-brane  with the value of the B-field 
in the spatial direction of the D2-brane is given by
\be
M_{D2} = \frac{1}{g\sqrt{2}} \sqrt{  1 + b^2}
\ee
This can be easily understood from the Dirac-Born-Infeld action of the
D2-brane. We have compactified the D2-brane on a square
torus of radius $\sqrt{2}$ for simplicity, the mass defect is
independent of this compactification.
The BPS mass formula of the bound state of the D0-brane
within the D2-brane is given by
\be
M^2 = \frac{1}{2g^2} (Q_0 + bQ_2)^2 + \frac{1}{2g^2} Q_2^2,
\ee
where $Q_0$ and $Q_2$ are the number of D0-branes and D2-branes
respectively.
Substituting the $Q_0=Q_2=1$ we obtain
\be
M_{D0-D2} = \frac{1}{g\sqrt{2}}\sqrt{ 1 + (1+b)^2}
\ee
The mass defect for the formation of the D0-D2 bound state is given by
\be
\Delta M = M_{D0-D2} - (M_{D0} + M_{D2})
\ee
We require the leading order term in the mass defect for
$b\rightarrow\infty$. This is given by
\be
\label{defect}
\Delta M = -\frac{1}{2\sqrt{2}gb^2}
\ee
Note that this is the mass defect obtained in \cite{agms} for the
formation of the D0-D2 bound state.
From the discussion in section 3.1 we note that the system reduces to
the D0-brane anti-D0-brane system for $b\rightarrow -\infty$. This
also reflected in the mass defect. We find that for $b\rightarrow
-\infty$ the mass defect $\Delta M = - \frac{\sqrt{2}}{g}$, which is
twice the mass of the D0-brane.

\section{The tachyon operators and their correlation functions}

In this section we construct the boundary operators $O_1$ and $O_2$
corresponding to the tachyon of the D0/D2 system. 
Then we write down the correlation functions of these operators needed
for the evaluation of the disc partition function.

\subsection{The tachyon operators}
The Hilbert space of the D0/D2 system can be
represented using a $2\times 2$ Matrix. 
\be
\left(
\begin{array}{cc}
(0,0)\hbox{Strings} \;\;&\;\; (0,2)\hbox{Strings} \\
(2,0)\hbox{Strings} \;\;& \;\; (2,2)\hbox{Strings} 
\end{array} \right)
\ee
The operator $O_1$ corresponding to the $(0,2)$ tachyon should involve
twist fields which change the boundary conditions. 
This is clear from
the fact that the $(0,2)$ strings have are fractionally moded.
We can see this in the mode expansions in \eq{bosemode1}, \eq{mode1}
and \eq{mode2}. Using these conditions the boundary operator $O_1$
is given by
\be
O_1 = \lambda_{+}\Sigma_\nu \sigma_+ + 
\lambda_{-}\Sigma_{-\nu} \sigma_-  ,
\ee
where
\label{twist1}
\be
\Sigma_\nu = \sigma_\nu e^{i\nu H} \;\;\; \Sigma_{-\nu} =
\sigma_{-\nu} e^{-i\nu H}
\ee
The twist operators $\sigma_\nu, \sigma_{-\nu}$ and $H$ are defined
in appendix A \eq{ope} and \eq{defh}. 
$\sigma_+$ and $\sigma_-$ are given by
\be
\sigma_+ = \left(
\begin{array}{cc}
0 & 1 \\
0 & 0
\end{array}
\right)
\;\;\;\;
\sigma_- = \left(
\begin{array}{cc}
0 & 0 \\
1 & 0
\end{array}
\right)
\ee
The operator $O_1$ does not have diagonal entries as expected.
The twist operator $\Sigma_\nu$ comes with the matrix $\sigma_+$ and
the anti-twist operator $\sigma_{-\nu}$ comes along with the matrix
$\sigma_-$. This is because the insertion of $\Sigma_{\nu}$ changes the
boundary conditions to that of the string joining the D0-brane to the
D2-brane. While the insertion of $\Sigma_{-\nu}$ changes the boundary
conditions to that of the string joining the D2-brane to the D0-brane.
that is a string of opposite orientation. $\lambda_+$ and $\lambda_-$
are complex conjugates of each other. The conformal dimension of the
operator $O_1$ is $\nu/2$

As $O_1$ is the bottom component of the the superfield,
the operator $O_2$ is obtained from $O_1$ by action of 
the supercurrent $G_{-1/2}$ on
the state $O_1$. This is done in appendix A in \eq{top1} and \eq{top2}.
We get 
\be
O_2 = \lambda_{+}\Lambda_\nu \sigma_+ + 
\lambda_{-}\Lambda_{-\nu} \sigma_-
\ee
where 
\be
\label{twist2}
\Lambda_\nu = -\frac{i}{\sqrt{2}} \tau_\nu e^{(\nu -1) H} \;\;\;\;
\Lambda_{-\nu} = -\frac{i}{\sqrt{2}} \tau_{-\nu} e^{-(\nu -1) H}
\ee
The excited twist operators $\tau_{\nu}$ and $\tau_{-\nu}$ are defined
in \eq{ope}. We note that 
the conformal dimension of $O_2$ is  $\nu/2+1/2$. The tachyon
superfield $\Gamma$ is a world sheet boson. This implies the $O_1$ is
a boson and $O_2$ is a world sheet fermion.

\subsection{The correlation functions}

To evaluate the disc partition function up to the the quartic order we
will need the two point and four point function of these operators.
We evaluate these correlators in the appendix. 
The two point functions of the twist operators is given by \footnote{
Here we have mapped the correlation functions found in appendix B and
appendix C to the boundary of the disc from the boundary of the upper
half plane.}
\be
\label{2pt1}
\langle \Sigma_\nu (\tau_1) \Sigma_{-\nu} (\tau_2)\rangle = \left| 2
\sin\frac{(\tau_1-\tau_2)}{2}\right|^{-\nu}  
\ee
This correlation function is fixed by the dimension of $\Sigma$. Its
normalization is fixed to be unity. The two point functions of the
excited twist operators $\Lambda$ is given by
\be
\label{2pt2}
\langle \Lambda_{\nu} (\tau_1) \Lambda_{-\nu} (\tau_2) \rangle
= -\nu \left|2
\sin\frac{(\tau_1-\tau_2)}{2}\right|^{-\nu -1} \Theta(\tau_1-\tau_2)
\ee
The normalization of this correlation function is determined in 
appendix B. 
In \eq{2pt1} and \eq{2pt2} we have written down the correlators on the
boundary of the unit disc and for arbitrary time orderings. The theta
function appears as the $\Lambda$'s are world sheet fermions.

The following four point functions are needed for the evaluation of
the disc partition function.
\be
\label{cor1}
\langle 
\Sigma_{-\nu} (\tau_1) \Sigma_{+\nu} (\tau_2) 
\Sigma_{-\nu} (\tau_3) \Sigma_{+\nu} (\tau_4) \rangle =
\tau_{12}^{-\nu} \tau_{13}^{+\nu} \tau_{14}^{-\nu}
\tau_{23}^{-\nu} \tau_{24}^{+\nu} \tau_{34}^{-\nu}
\frac{1}{F(\nu, 1-\nu, 1; x)}
\ee
where $x$ is the cross ratio given by
\be
x= \frac{\tau_{12}\tau_{34}}{\tau_{13}\tau_{24}}
\;\;\;\;
\hbox{and}\;\;\; \tau_{ij} = 2\sin\frac{(\tau_i - \tau_j)}{2}
\ee
$F(\nu, 1-\nu, 1;x)$ is the hypergeometric function defined in
\eq{hyper}. In \eq{cor1} we have evaluated the correlation function for
the time ordering $\tau_1>\tau_2>\tau_3>\tau_4$. We will see in 
section 5 
that we need the four point function in \eq{cor1} at $\nu =1$. At
$\nu=1$ the four point function simplifies. It is given by simple
Wick contractions.
\be
\label{cor1a}
\langle 
\Sigma_{-\nu} (\tau_1) \Sigma_{+\nu} (\tau_2) 
\Sigma_{-\nu} (\tau_3) \Sigma_{+\nu} (\tau_4) \rangle |_{\nu=1}=
|\tau_{12}\tau_{23}|^{-1} + |\tau_{14}\tau_{23}|^{-1}
\ee
In \eq{cor1a} we have written down the four point function at $\nu=1$
for arbitrary time orderings.

Now we need the four point function of the 
excited twist operator $\Lambda$.
This is given by
\be
\label{cor2}
\langle
\Lambda_{-\nu} (\tau_1)
\Lambda_{+\nu} (\tau_2)
\Lambda_{-\nu} (\tau_3)
\Lambda_{+\nu} (\tau_4) \rangle =
\tau_{12}^{-2h} \tau_{13}^{2h} \tau_{14}^{-2h}
\tau_{23}^{-2h} \tau_{24}^{2h} \tau_{34}^{-2h} G(x)
\ee
Here $h = \nu/2 + 1/2$ and  $G(x)$ is given \eq{defg}. 
The correlation function in \eq{cor2} is written down for the time
ordering $\tau_1>\tau_2>\tau_3>\tau_4$.
As before it 
is important to note that for $h=1$ the four point function in
\eq{cor2} simplifies to 
\bea
\label{cor2a}
&&\langle
\Lambda_{-\nu} (\tau_1)
\Lambda_{+\nu} (\tau_2)
\Lambda_{-\nu} (\tau_3)
\Lambda_{+\nu} (\tau_4) \rangle|_{\nu=1} = \\ \nonumber
&& 
\Theta(\tau_1-\tau_2) \Theta(\tau_3-\tau_4)
(\tau_{12}\tau_{34})^2 + 
\Theta(\tau_1-\tau_4) \Theta(\tau_2-\tau_3)
(\tau_{14}\tau_{23})^2
\eea
Here we have written the simplified four point function for $\nu=1$
and for arbitrary time orderings. The $\Theta$ function appears as the
$\Lambda$'s are world sheet fermions.

Finally we need the following four point function
\be
\label{cor3}
\langle
\Lambda_{-\nu} (\tau_1)
\Lambda_{+\nu} (\tau_2)
\Sigma_{-\nu} (\tau_3) \Sigma_{+\nu} (\tau_4) \rangle = 
\tau_{12}^{-2h} \tau_{23}^{\nu} \tau_{14}^{-\nu}
\tau_{23}^{-\nu}\tau_{24})^{\nu} \tau_{34}^{-\nu}
I(x)
\ee
Where $I(x)$ is given in \eq{defi} and the four point function in
\eq{cor3} is written down for $\tau_1>\tau_2>\tau_3>\tau_4$.
Now we need this correlation function with $\nu=1$.
For arbitrary time orderings this is given by
\be
\label{cor3a}
\langle
\Lambda_{-\nu} (\tau_1)
\Lambda_{+\nu} (\tau_2)
\Sigma_{-\nu} (\tau_3) \Sigma_{+\nu} (\tau_4) \rangle |_{\nu=1} = 
\Theta(\tau_3-\tau_4)
\tau_{12}^{-2}|\tau_{34}|^{-1}
\ee
Here again the $\Theta$ function appears because the $\Lambda$'s are
world sheet fermions. Notice that again for $\nu=1$ the four point
function in \eq{cor3} is determined by simple Wick contractions.

\section{Evaluation of the disc partition function}

In this section we evaluate the disc partition function using the
perturbative expansion in \eq{pert}. We will justify this expansion
for an almost marginal tachyon operator of the D0-D2 system.

\subsection{The quadratic term in the disc partition function}

The quadratic term $S_2$ in the disc partition function using the
renormalization prescription introduced by \cite{kmm2} is given by
\bea
\label{quad}
S_2 &=& -\langle S_{\rm{Boundary}}\rangle \\ \nonumber
&=& 
\lim_{\epsilon \rightarrow 0} -\frac{1}{4} 
\int \frac{d\tau}{2\pi}\langle
O_1(\tau) O_1(\tau-\epsilon) + O_2(\tau) \int \frac{1}{2} \Theta
(\tau -\tau'-\epsilon) O_2 (\tau') d\tau' \rangle.
\eea
This point splitting regularization was motivated by
the requirement of a finite one point function in \cite{kmm2}.
We now show that the divergences of the first term and the second term
in \eq{quad} cancel leading to a finite term. 
The first term in \eq{quad} is given by
\bea
S_2^1 &=& 
-\frac{1}{4}\int\frac{d\tau}{2\pi} \langle O_1(\tau) O_1(\tau
-\epsilon)\rangle \\ \nonumber
&=&
-\frac{1}{4}\int\frac{d\tau}{2\pi}\hbox{Tr} (\sigma_+\sigma_-) 
\langle\Sigma_+(\tau)\Sigma_-(\tau-\epsilon) \rangle
+ \hbox{Tr} (\sigma_-\sigma_+) 
\langle\Sigma_-(\tau)\Sigma_+(\tau-\epsilon) \rangle \\ \nonumber
&=& -\frac{\lambda_+\lambda_-}{2}  \int\frac{d\tau}{2\pi}
\frac{1}{(2\sin\frac{\epsilon}{2} )^\nu} \\ \nonumber
&=& -\frac{\lambda_+\lambda_-}{2} 
\frac{1}{(2\sin\frac{\epsilon}{2} )^\nu} 
\eea
Here we have used the two point function in \eq{2pt1}. For simplicity
we have taken $\epsilon$ to be positive. This can be done without
loss of generality.
Now the second term in \eq{quad} is given by
\bea
S_2^2 &=& -\frac{1}{4}\langle \int\frac{d\tau}{2\pi} O_2(\tau) \int
d\tau' \frac{\Theta(\tau -\tau'-\epsilon)}{2} O_2(\tau') \rangle \\
\nonumber
&=& \frac{\nu\lambda_+\lambda_-}{4}\int d\tau \frac{\Theta(\tau)
\Theta(\tau-\epsilon)}
{|2\sin\frac{\tau}{2}|^{2h}} \\ \nonumber
&=& \frac{\nu\lambda_+\lambda_-}{4}
\left( \int_0^{\pi} d\tau \frac{1}
{(2\sin\frac{\tau}{2})^{2h}} 
+ \int_0^\pi d\tau \frac{\Theta(\tau-\epsilon)}
{(2\sin\frac{\tau}{2})^{2h}}  \right)
\eea
Where we have taken the traces over the matrices. 
Substituting $u=\tan(\tau/2)$ and integrating by parts we obtain
\be
S_2^2 = \frac{\nu\lambda_+\lambda_- }{2^{2h+1}} \left( 
-\frac{4(1-h)}{2h-1} \int_0^\infty du 
\frac{u ^{2-2h} }{(1+u^2)^{2-h}}
+ \frac{2}{2h-1} \int_0^\infty\frac{du 
\delta(u-\tan\frac{\epsilon}{2})}
{u^{2h-1}(1+u^2)^{1-h} }
\right)
\ee
It is important to note that while integrating by parts 
the boundary terms either cancel or they are identically zero.
We expand in  $\epsilon$ and retain the singular and finite term in
$\epsilon$. We get
\be
S_2^2 = -(1-h)\frac{\pi}{4}\lambda_+\lambda_- + \frac{\nu}{2(2h-1)} 
\frac{1}{(2 \tan\frac{\epsilon}{2})^{2h-1})} \lambda_+\lambda_- 
\ee
We have retained the leading order in $(1-h)$ in the first term in the
above expression. Using  $2h-1 =\nu$ we see that 
on adding $S_2^1$ and $S_2^2$ 
see that the divergence in $\epsilon$ cancels and we obtain the
following finite term.
\be
S_2 = -(1-h)\frac{\pi}{4} \lambda_+\lambda_-
\ee

Note that the coefficient of the quadratic term is
proportional to the deviation from marginality of the boundary tachyon
interaction. Now the validity of the perturbative expansion in
\eq{pert} is clear. We are interested in  the stable 
minimum of the disc
partition function. The minimum can be evaluated to an accuracy of
$O((1-h)^2)$ if one knows the quartic term in the expansion in
\eq{pert}. For this it is sufficient to evaluate the quartic term with
$h=1$. This implies $\nu=1$. Furthermore the couplings at the minimum
are of the order of $O(1-h)$ which justifies the expansion for an
almost marginal operator.

\subsection{The quartic term in the disc 
partition function}

As we have mentioned above for the accuracy we are interested we can
evaluate the quartic term with $\nu =1$. Thus the four point functions
we will require are give in \eq{cor1a}, \eq{cor2a} and \eq{cor3a}.
The quartic term consists of three terms
\be
S_4 = \frac{1}{2} (S_4^1 + S_4^2 + S_4^3)
\ee
We will discuss them below.
$S_4^1$ is given by
\be
S_4^1 = \frac{1}{64\pi^2} \int d\tau d\tau' \langle
O_1(\tau)O_1(\tau-\epsilon) O_1(\tau')O_1(\tau'-\epsilon) \rangle
\ee
Substituting the four point function from \eq{cor1a}, 
taking the trace and using some
straight forward simplifications we obtain
\be
\frac{S_4^1}{(\lambda_+\lambda_-)^2} 
= \frac{1}{32\sin^2\frac{\epsilon}{2}} + \frac{1}{32\pi}
\int_0^\pi d\tau \frac{1}{|\sin \frac{\tau-\epsilon}{2}
\sin\frac{\tau+\epsilon}{2}|}
\ee
Simplifying further we obtain
\be
\label{q1}
\frac{S_4^1}{(\lambda_+\lambda_-)^2} 
=  \frac{1}{32\sin^2\frac{\epsilon}{2}} + \frac{1}{16\pi
\cos\frac{\epsilon}{2} \sin\frac{\epsilon}{2} }
\int_0^{\tan\frac{\epsilon}{2}} du
\left(\frac{1}{\tan\frac{\epsilon}{2} - u } +
\frac{1}{\tan\frac{\epsilon}{2} + u } \right)
\ee

Let us now focus on the next term $S_4^2$. It is given by
\bea
S_4^2 &=& \frac{1}{128\pi^2}\left(
\langle
\int d\tau_1 O_1(\tau_1) O_1(\tau_1 -\epsilon) \int d\tau_2 d\tau'
O_2(\tau_2) \Theta(\tau_2-\tau'-\epsilon) O_2(\tau')
\rangle \right. \\ \nonumber
&+& 
\left. \langle
\int d\tau_2 d\tau'
O_2(\tau_2) \Theta(\tau_2-\tau'-\epsilon) O_2(\tau')
\int d\tau_1 O_1(\tau_1) O_1(\tau_1 -\epsilon) 
\rangle
\right)
\eea
Using the four point  function in \eq{cor2a}, taking the traces
and performing straightforward manipulations we obtain
\bea
\label{q2}
\frac{S_4^2}{(\lambda_+\lambda_-)^2} &=& 
-\frac{1}{16}\frac{1}{\sin\frac{\epsilon}{2}}\int d\tau
\frac{\Theta(\tau) \Theta(\tau-\epsilon)}{(2\sin\frac{\epsilon}{2})^2}
\\ \nonumber
&=& -\frac{1}{16}
\frac{1}{\sin\frac{\epsilon}{2}\tan\frac{\epsilon}{2}}
\eea
Finally the third term $S_4^3$ is given by
\be
S_4^3
= \frac{1}{16}\langle \int \frac{d\tau_1}{2\pi} d\tau'
\frac{\Theta(\tau_1 -\tau' -\epsilon)}{2}
O_2(\tau_1) O_2(\tau')
\int \frac{d\tau_2}{2\pi} d\tau''
\frac{\Theta(\tau_2-\tau''-\epsilon)}{2}
O_2(\tau_2) O_2(\tau'')
\rangle
\ee
Substituting the correlation function of $O_2$ from \eq{cor3a} 
with $\nu=1$ and taking the
traces and using some straight forward manipulations we obtain
\bea
\label{q3}
\frac{S_4^3} {(\lambda_+\lambda_-)^2} 
&=&
\frac{1}{32} \frac{1}{(\tan\frac{\epsilon}{2})^2}
-\frac{1}{32\pi} 
\int_0^\pi d\tau \frac{1}{|\tan\frac{\tau +\epsilon}{2}
\tan\frac{\tau -\epsilon}{2} |}  \\ \nonumber
&=& 
\frac{1}{32} \frac{1}{(\tan\frac{\epsilon}{2})^2}
-\frac{1}{16\pi}\frac{1-\tan^2\frac{\epsilon}{2}}
{\tan\frac{\epsilon}{2}}
\int_0^{\tan\frac{\epsilon}{2}}
du \left(\frac{1}{\tan\frac{\epsilon}{2}-u} + 
\frac{1}{\tan\frac{\epsilon}{2}+u} \right) \\ \nonumber
&+& \frac{1}{32} 
- \frac{1}{8\pi}\epsilon
+\frac{1}{16\pi}\frac{1-\tan^2\frac{\epsilon}{2}}
{ \tan\frac{\epsilon}{2} }
\ln\left(\frac{1+\tan^2\frac{\epsilon}{2}}
{1-\tan^2\frac{\epsilon}{2}}\right)
\eea
Now taking 
the limit $\epsilon\rightarrow 0$ and adding the expressions in
\eq{q1}, \eq{q2} and \eq{q3} we find that all divergences cancel
leaving a finite term
\be
S_4 = \frac{1}{2} (S_4^1 + S_4^2 + S_4^3) =
\frac{1}{64} (\lambda_+\lambda_-)^2
\ee
Putting the quadratic and the quartic terms together we obtain the
following expansion for the disc partition function till the quartic
order
\bea
\frac{Z[\lambda_+\lambda_-]}{Z[0]} 
&=& 1 - \frac{(1-h)\pi}{4}\lambda_+\lambda_-
+ \frac{1}{64} (\lambda_+\lambda_-)^2 \\ \nonumber
&=&  1- \frac{1}{8b}\lambda_+\lambda_- +
\frac{1}{64} (\lambda_+\lambda_-)^2
\eea
We note that the above expansion of the disc partition function is
valid for the tachyon boundary operator which is almost marginal. We
have $h\sim 1$ for $b\rightarrow \infty$.
In the second line of the above equation we have substituted for $h$ 
in terms of $b$.

\section{Comparison with the mass defect}

According to the proposal of \cite{kmm2} 
the boundary string field theory 
action is expected to be the disc partition
function. Thus the difference in the 
value of the stable minimum and the unstable maximum 
of the disc partition
function is expected to reproduce the mass defect involved in tachyon
condensation.
For the case of the D0-D2 system the difference in the 
two critical points to $O((1-h)^2)$ is given by
\be
\label{stb}
Z[(\lambda_+\lambda_-)^*] -Z[0] = -\frac{Z[0]}{4b^2}
\ee
where $(\lambda_+\lambda_-)^* = 4/b$. $Z(0)$ is twice the mass of the
D0-brane. This can be seen as follows.
For $b\rightarrow -\infty$ we have seen that the system
effectively reduces to the D0-brane anti-D0-brane system. 
From \cite{kmm2} we see that the tachyon potential 
for the D0-brane anti-D0-brane is given by
\be
Z[0] e^{-\frac{(\lambda_+\lambda_-)^2}{4}}
\ee
where $Z[0]$ is twice the mass of the D0-brane. This reproduces the
right mass defect for the D0-brane anti-D0-brane system.
We can also understand the factor of $2$ from the trace of the
$2\times 2$ identity matrix in $Z[0]$.
Therefore the difference in the two critical points 
of the disc partition
function is given by
\be
Z[(\lambda_+\lambda_-)^*] - Z[0] = -\frac{1}{2\sqrt{2} gb^2} 
\ee
We have thus reproduced the 
leading order term for the mass defect 
in the formation of the bound
state of D0-D2 system given in \eq{defect} from the disc partition 
function.

\section{Conclusions}

We have evaluated the disc partition function 
for the D0-D2 system in the
presence of a large B-field using perturbation theory. We have shown
that the mass defect calculated from the disc partition function agrees
exactly with the expected result in the large B-field limit.
This lends support to the recent proposal \cite{kmm2} that
the disc partition function is the background independent superstring
field theory action. We can compare this result with that obtained
using Berkovits' string field theory. 
The tachyon potential for the D0-D4 system was evaluated at the zeroth
level of approximation in \cite{jus}. It is easy to extend that result
for the D0-D2 system. We see that the zeroth level truncation in
Berkovits' string field theory gives only $25\%$ of the expected
result for large value of the B-field. This implies that the tachyon
in the Berkovits' string field theory mixes with higher level fields.
So it seems that the disc partition function is better suited to study
the phenomenon of tachyon condensation. 

The disc partition function is perhaps undefined for arbitrary boundary
operators. In this paper we have shown that one can set up a valid
perturbative expansion of the disc partition function for boundary
operators which are almost marginal using the D0-D2 system as an
example.  We showed that the perturbative expansion is valid if one is
interested in the value of the disc partition at its
critical points. To obtain the mass defects in tachyon condensation the 
value of the critical points of the disc partition function is
sufficient.

In \cite{kmm2}
the disc partition function for a world sheet supersymmetric theory 
was also conjectured to be the boundary entropy \cite{aflud}. The disc
partition function satisfies the condition that it is stationary at
fixed points of the renormalization group flow and it takes the right
value. It would be interesting to see if it can be shown that the disc
partition function for the D0-D2 system decreases along the
renormalization group flow set up by the relevant tachyon operator.
This would be a tractable problem as the operator can be made almost
marginal
\footnote{It is easy to see that the disc partition function at the 
unstable maximum $\lambda_+= \lambda_- =0$ 
is higher than the stable minimum $(\lambda_+\lambda_-)^*$ from
\eq{stb}. The implies the renomalization group flow should drive the
system to the stable minimum.}.
This would provide additional evidence for the disc partition function
to coincide with the boundary entropy for a world sheet supersymmetric
theory.

\acknowledgments

The author thanks  K. Hori, K. Intrilligator, J. Kumar, B. Pioline 
and especially R. Gopakumar and S. Minwalla for discussions. He is
grateful for a discussion with E. Witten regarding the operator $O_2$.
He thanks the high energy theory group at Harvard for hospitality
where part of this work was done. The work of the author is supported
by NSF grant PHY97-2202.

\appendix
\section{The excited twist operators}
This appendix discuss the definition of the bosonic and fermionic
twist operators which are required to construct the tachyon operator
$O_1$ and $O_2$. 
The twist operators are located on the boundary of
the world sheet. Thus they are on the real axis when the world sheet
is the upper half plane. The insertion of the $\sigma_{\nu}$ changes
the boundary conditions of the string to that of a string joining the
D0-brane to the D2-brane. The anti-twist operator $\sigma_{-\nu}$
changes the boundary conditions to that of a string joining the
D2-brane to the D0-brane. From the mode expansions in
\eq{bosemode1} the bosonic twist operators $\sigma_\nu$ and
$\sigma_{-\nu}$ have the following OPE with the world sheet bosons
\bea
\label{ope}
\del X^+(z) \sigma_\nu(w) = \frac{1}{(z-w)^{(1-\nu) } } 
\tau_\nu (w) & \; &
\del X^-(z) \sigma_\nu(w) = \frac{1}{(z-w)^\nu} \tau^\prime_\nu (w) \\
\nonumber
\del X^+(z) \sigma_{-\nu}(w) = 
\frac{1}{(z-w)^{\nu} } \tau^\prime_{-\nu} (w) & \; &
\del X^-(z) \sigma_{-\nu}(w) = \frac{1}{(z-w)^{(1-\nu)} }
\tau_{-\nu} (w) \\
\nonumber
\bar{\del} X^+(\bar{z}) \sigma_\nu(w) = 
-\frac{1}{(\bar{z}-w)^{1-\nu}} \tau_\nu (w) & \; &
\bar{\del} X^-(\bar{z}) \sigma_\nu(w) = 
-\frac{1}{(\bar{z}-w)^\nu} \tau^\prime_\nu (w) \\
\nonumber
\bar{\del} X^+(\bar{z}) \sigma_{-\nu}(w) = 
-\frac{1}{(\bar{z}-w)^{\nu} } \tau^\prime_{-\nu} (w) & \; &
\bar{\del} X^-(\bar{z}) \sigma_{-\nu}(w) = 
-\frac{1}{(\bar{z}-w)^{(1-\nu)}} 
\tau_{-\nu} (w) \\
\nonumber
\eea
Here $w$ is a point on the real axis. The $\tau$'s are the excited
twist operators.
To construct the fermionic twist we first bosonize the fermions by
defining
\be
\label{defh}
\psi^+=i\sqrt{2}e^{iH} \;\;\; \psi^- = i\sqrt{2}e^{-iH}
\ee
where $H$ is a free boson.
The fermionic twist operator is given by $e^{i\nu H}$ and 
the anti-twist operator is given by $e^{-i\nu H}$.
The OPE  of these twist operators  
with the fermions is given by
\bea
\label{opef}
\psi^+(z) e^{i\nu H(w)} = i\sqrt{2}(z-w)^\nu e^{i(\nu+1)H(w)}
&\;& 
\psi^-(z) e^{i\nu H(w)} = \frac{i\sqrt{2}}{(z-w)^\nu} e^{-i(1-\nu)H(w)}
\\ \nonumber
\psi^+(z) e^{-i\nu H(w)} = \frac{i\sqrt{2}}{(z-w)^\nu} e^{i(1-\nu)H(w)}
&\;&
\psi^-(z) e^{-i\nu H(w)} = i\sqrt{2}(z-w)^\nu e^{-i(\nu+1)H(w)}
\eea
Now we have all the ingredients necessary 
for defining the twist operator
$\Sigma_\nu$ and $\Sigma_{-\nu}$ given in \eq{twist1}.
Using the OPE's  in \eq{ope} and \eq{opef} 
we find that the top component 
$\Lambda_\nu$ of the twist field
$\Sigma_\nu$ is given by  the following OPE
\be
\label{top1}
G(z)\Sigma_\nu(w) = \frac{1}{2} \frac{\Lambda_\nu(w)}{(z-w)}
\ee
Similarly $\Lambda_{-\nu}$ is given by
\be
\label{top2}
G(z)\Sigma_{-\nu}(w) = \frac{1}{2}
\frac{\Lambda_{-\nu}}{(z-w)}
\ee
Here $G(z)$ is given by
\be
G(z) = -\frac{1}{4} (\psi^+\del X^- + \psi^-\del X^+) 
\ee
Here we have defined the supercurrent $G(z)$ only for the relevant two
coordinates, $X^+$ and $X^-$.
Thus the top component of the twist field is given by
\be
\Lambda_\nu = -\frac{i}{\sqrt{2}} \tau_\nu e^{-i(1-\nu)H}
\;\;\;\;\;\;
\Lambda_{-\nu} = -\frac{i}{\sqrt{2}} \tau_\nu e^{i(1-\nu)H}
\ee

\section{Normalization of the two point function the twist
operators}
In this section we discuss the normalization of the two point function
given in \eq{2pt1} and \eq{2pt2}. 
We first normalize the two point function of the bosonic twist
operator to be
\be
\label{s2pt}
\langle \sigma_\nu(z) \Sigma_{-\nu} (w)  \rangle
= \frac{1}{(z-w)^{\nu(1-\nu)}}
\ee
This normalization fixes the two point function of the twist operator 
$\Sigma_{\nu}$ to be
\be
\langle \Sigma_\nu(z) \Sigma_{-\nu} (w) \rangle
= \frac{1}{(z-w)^{\nu}}
\ee

Now we fix the normalization of the two point function of the excited
twist operators.
Consider the correlator
\be
g(z,w) = \frac{\langle -\frac{1}{2} \del X^+(z) \del X^-(w) 
\sigma_{-\nu}(z_1) \sigma_{\nu}(z_2) \sigma_{-\nu}(z_3)
\sigma_{\nu}(z_4) \rangle}
{\langle \sigma_{-\nu}(z_1) \sigma_{\nu}(z_2)
\sigma_{-\nu}(z_3)
\sigma_{\nu}(z_4) \rangle}
\ee
Taking the limit
\be
\label{l4pt}
\lim_{
z \rightarrow z_2 w\rightarrow z_1 } 
\left( (z-z_2)^{(1-\nu)} (w-z_1)^{(1-\nu)} g(z,w) \right) =
-\frac{1}{2}\frac{\langle \tau_{-\nu}(z_1) \tau_\nu (z_2)
\sigma_{-\nu}(z_3) \sigma_{\nu}(z_4) \rangle}{\langle
\sigma_{-\nu}(z_1) \sigma_{\nu}(z_2) \sigma_{-\nu}(z_3)
\sigma_{\nu}(z_4) \rangle}
\ee
we can obtain the four point function
$\langle \tau_{-\nu}(z_1) \tau_\nu (z_2)
\sigma_{-\nu}(z_3) \sigma_{\nu}(z_4) \rangle$. From this we will
determine the normalization of the two point functions of
$\sigma_\nu$.
Evaluation of $g(z,w)$ was reviewed and 
$\langle \sigma_{-\nu}(z_1) \sigma_{\nu}(z_2) \sigma_{-\nu}(z_3)
\sigma_{\nu}(z_4) \rangle$
in \cite{jus}. Here we just state
the result.
\bea
g(z, w) &=& \omega_\nu(z) \omega_{1-\nu} (w)  
\left(
\nu\frac{(z-z_1)(z-z_3)(w-z_2)(w-z_4)}{(z-w)^2} \right.\\ \nonumber
&+& \left. (1-\nu) \frac{(z-z_2)(z-z_4)(w-z_1)(w-z_3)}{(z-w)^2} +
A(z_1,z_2,z_3, z_4) \right)
\eea
Where $\omega_{\nu} (z)$ is given by
\be
\omega_\nu(z) =
\frac{1}{[(z-z_1)(z-z_3)]^\nu}\frac{1}{[(z-z_2)(z-z_4)]^{1-\nu}}
\ee
and 
\be
A(0, x, 1, z_4)= -z_4 x(1-x) \frac{d}{dx}\ln F(\nu, 1-\nu , 1 ; x)
\ee
here $z_4$ stands for $\infty$. $F(\nu, 1-\nu, 1;x)$ is the
hypergeometric function whose integral representation is given by
\be
\label{hyper}
F(\nu, 1-\nu,1; x) = \frac{1}{\pi} \sin(\pi\nu) \int_0^1 dy
\frac{1}{ y^\nu (1-y)^{1-\nu} (1-xy)^\nu }
\ee
The four point function of the bosonic twists are given by
\be
\label{s4pt}
\langle \sigma_{-\nu}(z_1) \sigma_{\nu}(z_2) \sigma_{-\nu}(z_3)
\sigma_{\nu}(z_4) \rangle= 
z_{12}^{-2h'}
z_{13}^{2h'}
z_{14}^{-2h'}
z_{23}^{-2h'}
z_{24}^{2h'}
z_{34}^{-2h'}\frac{1}{F(\nu, 1-\nu, 1;x)}
\ee
where $z_{ij} = z_i-z_j$, $h' = \nu(1-\nu)/2$  
and $x = (z_{12} z_{34})/(z_{13}z_{24})$.
Using \eq{l4pt} and \eq{s4pt} we find 
\be
\label{t4pt}
\langle
\tau_{-\nu} (0) \tau_\nu(x) \sigma_{-\nu}(1)
\sigma_{\nu}(z_4) \rangle =
\frac{1}{ x^{\nu(3-\nu)} (1-x)^{\nu(2-\nu)} z_4^{\nu(1-\nu)} }
I(x)
\ee
where $I(x)$ is given by
\be
\label{defi}
I(x) =
\frac{1}{F(\nu,1-\nu, 1; x)}
\left( \nu (1-x) + x (1-x)\frac{d}{dx} \ln F(\nu, 1-\nu, 1;x) \right)
\ee
It is easy to see that using \eq{s2pt} and \eq{t4pt}
we obtain the following normalization for the two point functions of the
excited twist operators.
\be
\langle \tau_{-\nu} (z_1) \tau_\nu (z_2) \rangle
= \frac{2\nu}{ (z_1-z_2)^{ \nu(3-\nu) } }
\ee
Using the above equation we see that 
\be
\langle \Lambda_{\nu} (z_1) \Lambda_{-\nu} (z_2) 
\rangle = - \frac{\nu}{(z_1-z_2)^{\nu +1}}
\ee

\section{The four point function of the excited twist operators}

In this section we will focus on the evaluation of the 
four point function involving the excited twist operators.
We first evaluate the four point function of the excited twist
operators given by
$\langle \tau_{-\nu}(z_1) \tau_{\nu}(z_2) \tau_{-\nu}(z_3)
\tau_{\nu}(z_4)\rangle$.
To do this first consider the auxiliary correlator
\be
f(z, w, z', w') = 
\frac{ \langle 
E(z,w,z',w')
\sigma_{-\nu} (z_1) \sigma_{\nu} (z_2) \sigma_{-\nu}(z_3)
\sigma_{-\nu} (z_4) \rangle}
{ \langle \sigma_{-\nu} (z_1) \sigma_{\nu} (z_2) \sigma_{-\nu}(z_3)
\sigma_{-\nu} (z_4) \rangle }
\ee
where $E(z,w,z',w')$ is given by
\be
E(z,w,z,w')= 
\left(-\frac{1}{2}\del X^+ (z) \del X(w) \right)
\left( -\frac{1}{2} \del X^+(z') \del X(w') \right)
\ee
The function $f(z, w, z', w')$ can be determined by singularity
structure and monodromy conditions.
The following form for $f(z,w,z',w')$ has the required 
singularity structure. 
\bea
f(z,w,z',w') &=&
\omega_{\nu} (z)
\omega_{1-\nu} (w)
\omega_{\nu} (z')
\omega_{1-\nu} (w')\\ \nonumber
&\times& \left[
e(z,w) e(z',w') + e(z',w)e(z,w') + B(z_1, z_2, z_3, z_4)
\right]
\eea
The function $e(z,w)$ is given by
\bea
e(z,w)&=& 
\left(
\nu\frac{(z-z_1)(z-z_3)(w-z_2)(w-z_4)}{(z-w)^2} \right.
\\ \nonumber
&+& \left. (1-\nu) \frac{(z-z_2)(z-z_4)(w-z_1)(w-z_3)}{(z-w)^2} +
A(z_1,z_2,z_3, z_4) \right)
\eea
Here $B(z_1,z_2,z_3,z_4)$ is a function which can be fixed by
monodromy conditions. That is, the change in $X^+$ from going to one
Dirichlet boundary to another is zero. Using this input we find that
$B(z_1, z_2, z_3, z_4)$ is zero.
Now taking the limit
\be
\lim_{z\rightarrow z_2, w\rightarrow z_1, z'\rightarrow z_4,
w'\rightarrow z_3}
\left( (z-z_2)(w-z_1)(z'-z_4)(w'-z_3)\right)^{1-\nu} f(z, w, z', w')
\ee
we can extract out the required correlation function.
We find
\be
\langle \tau_{-\nu}(z_1) \tau_{\nu}(z_2) \tau_{-\nu}(z_3)
\tau_{\nu}(z_4) \rangle
= 
\frac{4}{
z_{12}^{\nu(3-\nu)}
z_{13}^{-\nu(3-\nu)}
z_{14}^{\nu(3-\nu)}
z_{23}^{\nu(3-\nu)}
z_{24}^{-\nu(3-\nu)}
z_{34}^{\nu(3-\nu)} }
G(x)
\ee
where $G(x)$ is given by
\bea
\label{defg}
G(x) &=& \frac{1}{F(\nu, 1-\nu, 1, x)} \left(
(1-x)^2
\left[ \nu + x\frac{d}{dx}\ln F(\nu, 1-\nu, 1, x) \right]^2 \right. 
\\ \nonumber
&+&
\left. x^2\left[ 
\nu + (1-x)\frac{d}{dx}\ln F(\nu, 1-\nu, 1, x) \right]^2
\right)
\eea
Putting the bosonic and the fermionic twists together one obtains
\be
\langle
\Lambda_{-\nu} (z_1) \Lambda_{\nu} (z_2) \Lambda_{-\nu} (z_3)
\Lambda_{\nu} (z_4) \rangle
=
z_{12}^{-(\nu +1)}
z_{13}^{\nu +1}
z_{14}^{-(\nu +1)}
z_{23}^{-(\nu +1)}
z_{24}^{(\nu +1)}
z_{34}^{-(\nu +1)}
G(x)
\ee
Now using \eq{t4pt} and the definition of the $\Lambda$ twist operators
we find
\be
\langle
\Lambda_{-\nu}(z_1) \Lambda_{\nu}(z_2) \Lambda_{-\nu}(z_3)
\Lambda_{\nu}(z_4) \rangle =
z_{12}^{-2h} z_{l3}^{\nu} z_{l4}^{-\nu} z_{23}^{-\nu}
z_{24}^{\nu} z_{34}^{-\nu}I(x)
\ee
This completes the derivation of the required four point functions.

\end{document}